\documentclass[conference]{IEEEtran}
\usepackage{comment}

\usepackage{amsthm,amsmath,amssymb}
\usepackage{mathrsfs}

\usepackage{cite}

\usepackage{amsmath,amssymb,amsfonts}
\usepackage{algorithmic}
\usepackage{graphicx}
\usepackage{textcomp}

\usepackage{cite}

\usepackage{amsmath,amssymb,amsfonts}
\usepackage{algorithmic}
\usepackage{graphicx}
\usepackage{epstopdf}

\usepackage{subfigure}
 \graphicspath{{\figures}}
\usepackage{textcomp}
\usepackage{xcolor}
\usepackage{verbatim}
\usepackage{makecell}
\usepackage{booktabs}
\usepackage{CJK}
\usepackage{multirow}

\usepackage{algorithm}
\usepackage{algorithmic}

\makeatletter
\newcommand{\removelatexerror}{\let\@latex@error\@gobble}
\makeatother

\IEEEoverridecommandlockouts   % show \thanks
\begin{document}

\begin{CJK}{UTF8}{gbsn}

\title{Adaptive Bitrate Video Semantic Communication over Wireless Networks}
% \author{Haonan Tong, Ye Hu, Sihua Wang, Changchuan Yin\\
        % hntong@bupt.edu.cn; yeh17@vt.edu; sihuawang@bupt.edu.cn; ccyin@bupt.edu.cn}

\begin{comment}
\author{Haonan Tong\IEEEauthorrefmark{1}, Zhaohui Yang\IEEEauthorrefmark{2}, Sihua Wang\IEEEauthorrefmark{1}, Changchuan Yin\IEEEauthorrefmark{1}\\
 \IEEEauthorrefmark{1}  
Beijing Key Laboratory of Network System Architecture and Convergence,
Beijing University of Posts and Telecommunications, Beijing, China 100876\\
Emails: \{hntong,sihuawang,ccyin\}@bupt.edu.cn.\\
\IEEEauthorrefmark{2} 
Department of Electronic and Electrical Engineering, University College London, WC1E 6BT London, UK.
Centre for Telecommunications
Research, Department of Engineering, King's College London, WC2R 2LS,
UK\\
Email: yang.zhaohui@kcl.ac.uk.
}
\end{comment}  

% \author{Haonan~Tong,~\IEEEmembership{Member,~IEEE},~Zhaohui Yang,~\IEEEmembership{Member,~IEEE},~Sihua Wang,~\IEEEmembership{Member,~IEEE},\\
% ~and Changchuan Yin,~\IEEEmembership{Senior Member,~IEEE}

% \author{Wentao Gong\IEEEauthorrefmark{1}, Haonan Tong\IEEEauthorrefmark{1}, Sihua Wang\IEEEauthorrefmark{1}, and Changchuan Yin\IEEEauthorrefmark{1}\\
\author{Wentao Gong\IEEEauthorrefmark{1}, Haonan Tong\IEEEauthorrefmark{1}, Sihua Wang\IEEEauthorrefmark{1}, Zhaohui Yang\IEEEauthorrefmark{2}, Xinxin He\IEEEauthorrefmark{1}, and Changchuan Yin\IEEEauthorrefmark{1}\\
% \small \IEEEauthorrefmark{1}  
Beijing Key Laboratory of Network System Architecture and Convergence\\
Beijing University of Posts and Telecommunications, Beijing, China \\
\IEEEauthorrefmark{2} College of Information Science and Electronic Engineering, Zhejiang University, Hangzhou, China\\
% \IEEEauthorrefmark{2} 
% Department of Electronic and Electrical Engineering, University College London, WC1E 6BT London, UK.\\
% \IEEEauthorrefmark{3}
% Wireless@VT, Bradley Department of Electrical and Computer Engineering, Virginia Tech, Blacksburg, VA, USA.\\
Emails: \{gongwentao, hntong, sihuawang, hxx\_9000, ccyin\}@bupt.edu.cn, yang\_zhaohui@zju.edu.cn
\vspace{-0.5cm}
% \thanks{
% }
}

\maketitle
\vspace{-2cm}

\begin{abstract}
This paper investigates the adaptive bitrate (ABR) video semantic communication over wireless networks. In the considered model, video sensing devices must transmit video semantic information to an edge server, to facilitate ubiquitous video sensing services such as road environment monitoring at the edge server in autonomous driving scenario. However, due to the varying wireless network conditions, it is challenging to guarantee both low transmission delay and high semantic accuracy at the same time if devices continuously transmit a fixed bitrate video semantic information. To address this challenge, we develop an adaptive bitrate video semantic communication (ABRVSC) system, in which devices adaptively adjust the bitrate of video semantic information according to network conditions. Specifically, we first define the quality of experience (QoE) for video semantic communication. Subsequently, a swin transformer-based semantic codec is proposed to extract semantic information with considering the influence of QoE. Then, we propose an Actor-Critic based ABR algorithm for the semantic codec to enhance the robustness of the proposed ABRVSC scheme against network variations. Simulation results demonstrate that at low bitrates, the mean intersection over union (MIoU) of the proposed ABRVSC scheme is nearly twice that of the traditional scheme. Moreover, the proposed ABRVSC scheme, which increases the QoE in video semantic communication by 36.57\%, exhibits more robustness against network variations compared to both the fixed bitrate schemes and traditional ABR schemes.

% 研究了视频语义通信系统的自适应比特率问题。在所考虑的系统模型中，视觉传感器必须将视频语义分割发送到边缘服务器，以支持面向目标的应用，如视频会议等。由于传输信道的波动，如果传感器总是传输高比特率的视觉语义信息，语义通信可能无法保证低传输延迟。为了解决这一问题，我们提出了一种自适应比特率视频语义通信(abvsc)系统，该系统中视觉传感器根据网络条件自适应地调整视觉语义信息的比特率。特别地，我们首先对面向目标的视频语义通信的体验质量(QoE)进行建模。然后，为了优化QoE，我们提出了一种基于Swin-Transformer的语义编解码器来提取语义信息。接下来，我们将基于Actor-Critic的ABR算法引入到语义编解码器中，从而提高了所提出的abvsc的鲁棒性。仿真结果表明，在保持mIoU的情况下，ABRVSC比传统编码方案可有效减少50%的传输数据量。此外，所提出的ABRVSC比基线能更好地适应网络动态，并减少5%的视频下载延迟。
% This paper proposes a bitrate adaptive video semantic segmentation transmission system. The system realizes semantic communication by semantic segmentation, encoding and decoding of different scenes in the video. At the same time, a video rate adaptive algorithm based on video semantic segmentation coding is proposed, which can automatically adjust the rate and semantic compression ratio according to the dynamic changes of the network and improve the efficiency of the system. Experiments show that our system makes video semantic information better adapt to network dynamic changes, and achieves higher QoE and lower communication download delay. At the same time, our system can extract semantic features more effectively, and reduce the amount of data while maintaining mIoU.

\end{abstract}

\section{Introduction}
% Emerging network applications (e.g., digital twin and metaverse) in future sixth-generation (6G) wireless communications necessitate ubiquitous perception which poses significant challenges for resource-limited wireless networks. Consequently, efficient transmission is crucially required by emerging applications. 

% In goal-oriented communication applications (e.g., obstacle recognition in autonomous driving and human body recognition in video interaction services), video transmitters only need to convey the semantically aware behaviors of the primary objects, rather than all the details of the objects. As a result, with achieving the goal of communication using fewer data transmissions, video semantic communication based on segmentation is a powerful enabler for ubiquitous perception.
Future wireless network is required to support Internet of everything (IoE) which provides ultimate user experiences through ubiquitous connectivity and sensing techniques. In practical scenarios such as smart home and autonomous driving, wireless devices must transmit massive amounts of data (e.g., texts, audios, videos) to achieve ubiquitous sensing. However, due to the huge data amount of videos, it is a large overhead for wireless networks with limited spectrum resources to support pervasive video transmission. Thus, this motivates the implementation of video semantic communication techniques which can reduce the transmitted data amount through transmitting small-size video semantic information. For instance, in autonomous driving scenario, vehicles primarily concentrate on the locations of pedestrians and buildings for obstacle avoidance, rather than all the pixels sampled by vehicle cameras. In such applications, by substituting raw data transmission with obstacle-related semantic information, semantic communication techniques can reduce the bandwidth occupation of video transmission while maintaining the accuracy of task-oriented communication. However, the adaptive scheme for video semantic communication over dynamic wireless networks has not been well designed. 

Recent works \cite{farsad2018deep,huang2021deep,tong2021federated,xie2022task} have investigated the efficiency and robustness of semantic communication.
In \cite{farsad2018deep}, an efficient text codec was proposed to achieve lower word error rate with high transmission efficiency. Considering the image data, the authors in \cite{huang2021deep} employed a generative adversarial networks (GANs)-based image semantic codec to enhance the semantic consistency of the received images which reduced the amount of the transmitted data. Besides, in \cite{tong2021federated}, an autoencoder was proposed for audio semantic communication to improve the spectrum efficiency while maintaining accuracy. Furthermore, the authors in \cite{xie2022task} proposed an efficient semantic communication system for the multimodal multi-user scenarios which improved visual question answering (VQA) accuracy under low signal to noise ratio (SNR) conditions. However, focusing on the accurate semantic extraction, existing works in \cite{farsad2018deep,huang2021deep,tong2021federated,xie2022task} have not considered the delay guarantee problem for semantic communication, especially with varying transmission environments.

The prior contributions \cite{zhou2021semantic,zhou2022adaptive,huang2022toward} have explored adaptive mechanisms for semantic communication in varying transmission environments. In \cite{zhou2021semantic}, an adaptive transformer based codec was introduced to the text semantic communication system, enabling the system to flexibly adapt to channel variations. In \cite{zhou2022adaptive}, the authors proposed an adaptive bit-length scheme for text semantic communication to reduce the transmitted bit amount avoiding a significant accuracy decrease. Furthermore, in \cite{huang2022toward}, the authors improved the results of the reconstructed image through an adaptive coding method that assigns different numbers of bits to different semantic important levels. However, existing studies \cite{zhou2021semantic,zhou2022adaptive,huang2022toward} have not considered an adaptive video bitrate mechanism tailored for video semantic transmission. 
Given that video semantic communication has a stronger demand for delay guarantee compared to other adaptive systems, it calls for an adaptive bitrate (ABR) mechanism that adapts to varying network conditions while meeting the requirements of high semantic accuracy and low delay.

To address this issue, in this paper, we propose an adaptive bitrate video semantic communication (ABRVSC) system to adapt the bitrate of video semantic information to the network variation. The main contributions of this paper are as follows:

1) We develop an ABRVSC system in which devices adaptively adjust the bitrate of video semantic information according to network conditions. 2) We first define the quality of experience (QoE) for video semantic communication and formulate a QoE maximization problem to improve the semantic accuracy while meet the delay requirements of video semantic communication. 3) To solve the proposed problem, we then propose a swin transformer-based semantic codec for extracting semantic information, and further, introduce an Actor-Critic based ABR algorithm for the semantic codec to select the appropriate bitrate and to guarantee low transmission delay and high semantic accuracy. 4) Simulation results with CamVid dataset demonstrate that at low bitrates, the mean intersection over union (MIoU) of the proposed ABRVSC scheme is nearly twice that of the traditional scheme. Moreover, the proposed ABRVSC scheme, which improves the QoE in video semantic communication by 36.57\%, exhibits more robustness against network variations compared to both the fixed bitrate schemes and traditional ABR schemes.

The rest of this paper is organized as follows. Section II introduces the system model and problem formulation. Section III provides detailed descriptions of the proposed codec and Actor-Critic based video ABR algorithm. Simulation results are presented in Section IV. Conclusion is drawn in Section V.

\begin{figure}[t]
\centering
\includegraphics[width=0.9\linewidth,height=0.6\linewidth]{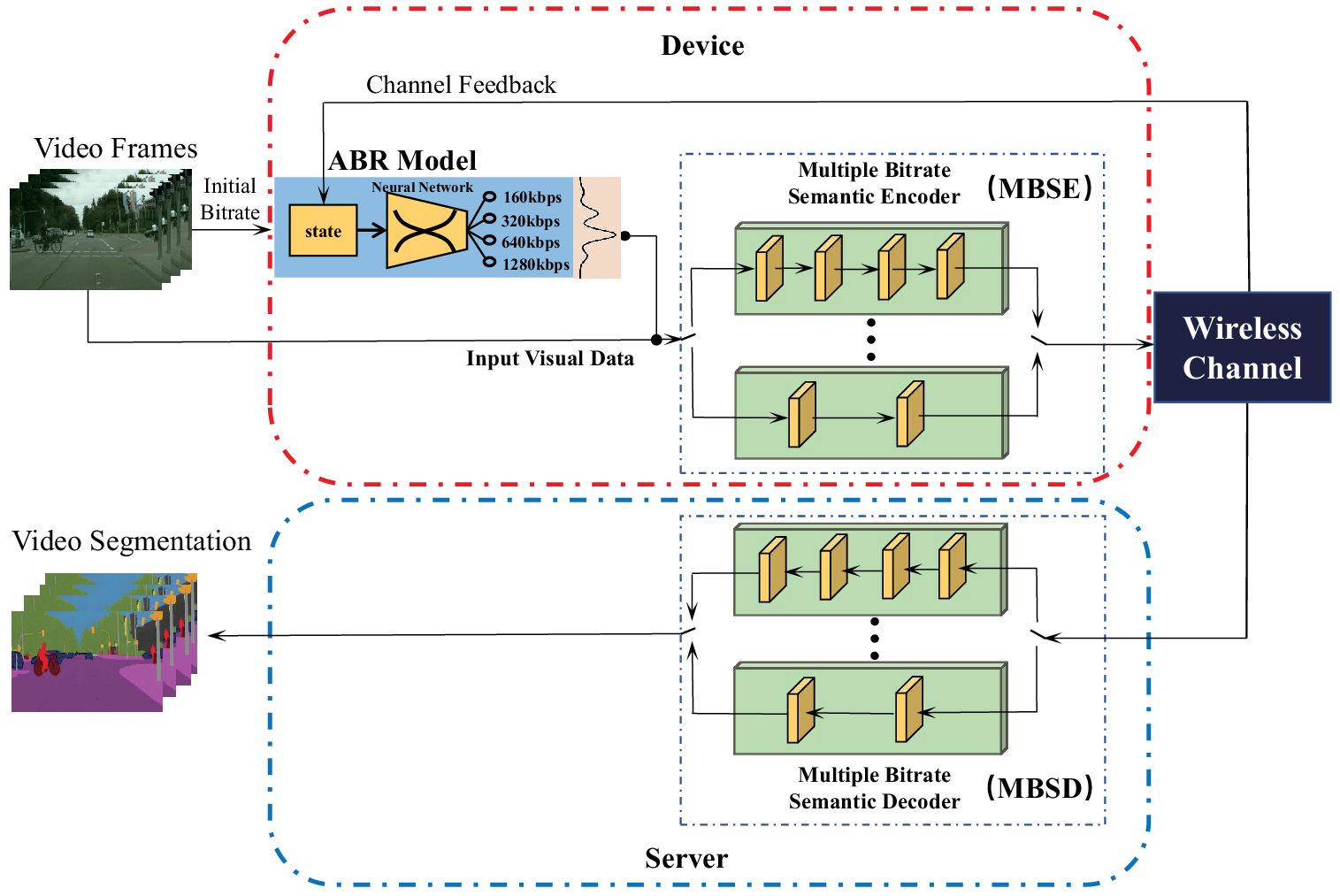}
\setlength{\abovecaptionskip}{-0.1cm}
\caption{The proposed ABRVSC system.}
\label{fig1:system_model}
\vspace{-0.4cm}
\end{figure}

\section{System Model}

We consider an ABRVSC system consisting of a device and a server, as shown in Fig.~\ref{fig1:system_model}. In the considered system, the device (e.g., vehicle) must transmit video semantic information which contains locations of key objects (e.g, humans and vehicles) in the sensing area, to an edge server. Then the edge server utilizes the information to sense the road environment, thus enhancing the server's ability to manage the road situation. To efficiently achieve this goal of video data transmission, the device must extract the accurate video semantic segmentation features using a semantic encoder and the server must reconstruct the video segmentation using a semantic decoder. Meanwhile, due to the dynamic characteristics of wireless network environments, the bitrate of transmitted information cannot be always guaranteed during a long time when the channel seriously degrades. As a result, the video semantic segmentation features must be encoded into multiple bitrates to guarantee both low transmission delay and high semantic accuracy by adjusting the bitrate when network condition varies. To meet the aforementioned requirements, an ABR module and a multiple bitrate semantic encoder (MBSE) module are deployed at the device. The ABR module first selects the appropriate bitrate, and then the MBSE encodes the video into semantic features with the determined bitrate. Given that the selection of bitrate precedes the encoding process, the computational complexity of our proposed system aligns with that of systems using a fixed bitrate. As such, the proposed scheme is suitable for deployment on the device for video semantic encoding. On the server side, the received semantic features, which have passed through the wireless channel, are processed by a deployed multiple bitrate semantic decoder (MBSD) module to reconstruct the required video segmentation.

% The device incorporates an Adaptive Bitrate (ABR) module and a Multiple Bitrate Semantic Encoder (MBSE). The server side deploys a Multiple Bitrate Semantic Decoder (MBSD) module. The complete ABRVSC system is shown in Fig.~\ref{fig1:system_model}. In this system, the device transmits video data, which primarily comprises semantic-aware information of key objects such as humans, vehicles, and obstacles, to the server. As the server-side application mainly focuses on these essential components, the device is only required to transmit semantic segmentations. 
% To achieve accurate extraction of semantic information on the device side, the MBSE module is designed to extract relevant semantic features and generate multi-bitrate video semantics. In order to ensure low transmission delay, the ABR module dynamically selects the appropriate bitrate based on the network conditions. On the server side, the MBSD module is responsible for reconstructing information from the received semantic features, ultimately achieving the specific goal of video transmission. 

In this section, we first present the details of the video semantic transmission model in the developed ABRVSC scheme. Subsequently, we introduce the metric for semantic accuracy. Finally, we propose a QoE model for the ABRVSC  scheme and formulate the QoE maximization problem.

\subsection{Video Semantic Transmission Model}
We assume that a device collects videos using a visual sensor (e.g., a camera) and preprocesses the video data (denoising and enhancing) to generate the raw video $\boldsymbol{v}$ with an initial bitrate $B$. The raw video $\boldsymbol{v}$ consists of $n$ video chunks $\left\{ \boldsymbol{v}_{1},\boldsymbol{v}_{2},...,\boldsymbol{v}_{n} \right\}$. Before transmission, the device must select an appropriate video bitrate $b_n \in \mathcal{B}$ for each chunk $\boldsymbol{v}_{n}$ according to the ABR algorithm, to adapt to the network conditions, where $\mathcal{B}$ is the set of bitrates available for selection. Then, the bitrate $b_n$ and video chunk $\boldsymbol{v}_{n}$ are delivered to the MBSE module for semantic extraction and coding.

Due to the varying network conditions, it is challenging to guarantee both low transmission delay and high semantic accuracy at the same time if the device continuously transmit fixed bitrate video semantic information. Therefore, it is necessary to pretrain MBSE models with different compression ratios offline. The MBSE module selects the compression ratio according to the bitrate and video chunk for semantic extraction encoding. The bitrate of video chunk $\boldsymbol{v}_{n}$ can be given by:
\begin{equation} 
    {{b}_{n}}=\frac{B}{{{c}_{n}}},
\end{equation} 
where ${{c}_{n}}$ is the compression ratio of the video chunk $\boldsymbol{v}_{n}$. 

Given the video chunk $\boldsymbol{v}_{n}$, semantic feature of 
$\boldsymbol{v}_{n}$ needs to be extracted before transmission.
Let ${{E}_{n}}(\cdot, b_n )$ be the encoding function of the MBSE module which encodes video with bitrate $b_n$. The relationship between the semantic feature $\boldsymbol{x}_{n}$ and input $\boldsymbol{v}_{n}$ can be given by:
\begin{equation}
    {\boldsymbol{x}_{n}}={{E}_{n}}(\boldsymbol{v}_{n}, b_n).
\end{equation}

The encoded semantic feature is transmitted over wireless channels and the received semantic feature at the receiver is given by:
\begin{equation}
    {\boldsymbol{y}_{n}}=h \boldsymbol{x}_{n}+ \boldsymbol{z},
\end{equation}
where $\boldsymbol{y}_{n}$ is the semantic feature received by the semantic decoder, $h$ is the Rayleigh fading channel coefficient, and $\boldsymbol{z}$ is the additive noise which follows a Gaussian distribution, $\boldsymbol{z} \sim$ $\mathcal{N}\left(0, \sigma^2 \boldsymbol{I}\right)$, with $\sigma^2$ being variance and $\boldsymbol{I}$ being the identity matrix.

The server selects the corresponding decoder model to decode the received semantic feature $\boldsymbol{y}_{n}$. The function of the MBSD module is represented by ${{D}_{n}}(\cdot, b_n)$. Thus, the correlation between the output video segmentation $\boldsymbol{o}_{n}$ and the received semantic features $\boldsymbol{y}_{n}$ is given by:
\begin{equation}
    \boldsymbol{o}_{n}={{D}_{n}}(\boldsymbol{y}_{n},b_n).
\end{equation}

\subsection{The Metric for Semantic Accuracy} 
A video chunk consists of a continuous sequence of frames. In the ABRVSC scheme, the objective of semantic segmentation is to assign a label to each pixel in a frame, denoting the object or region to which it belongs. This process is driven by the primary objective of enhancing the average accuracy of model-based segmentation across all frames. Consequently, semantic segmentation tasks utilize MIoU as the accuracy metric, which is calculated by:
\begin{equation}
    MIoU=\frac{1}{c}\sum\limits_{i=0}^{c-1}{\frac{{{p}_{ii}}}{\sum\limits_{j=0}^{c-1}{{{p}_{ij}}+\sum\limits_{j=0}^{c-1}{{{p}_{ji}}-{{p}_{ii}}}}}},
\end{equation}
where $c$ is the total number of categories and ${{p}_{ij}}$ is the count of pixels in $\boldsymbol{o}_{n}$ predicted as category $j$, which actually belong to category $i$. MIoU assesses the segmentation accuracy of the model for all categories and is given an average value, ranging from 0 to 1. A higher MIoU means a more accurate segmentation.

\begin{figure}[t]
\centering
\includegraphics[width=0.95\linewidth]{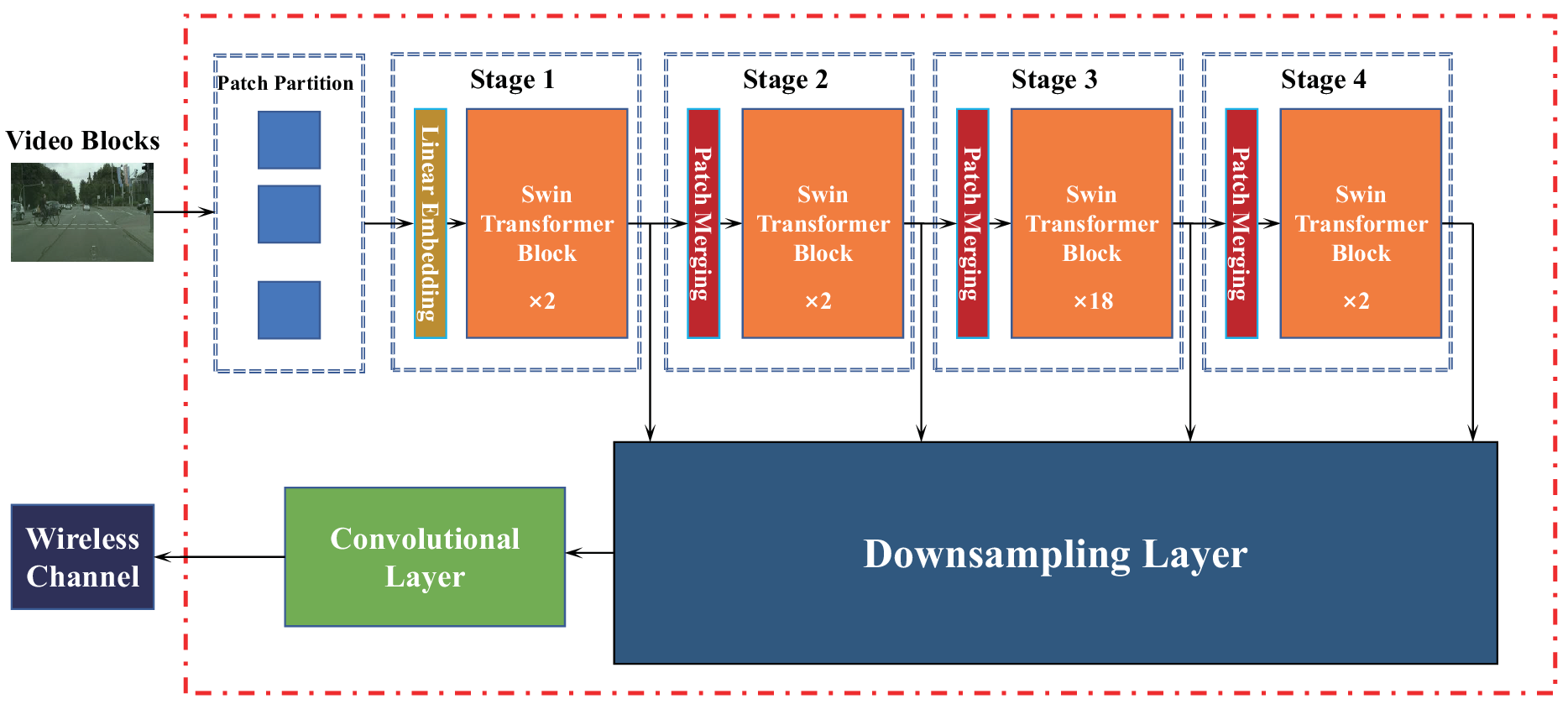}
\setlength{\abovecaptionskip}{-0cm}
\caption{The framework of swin transmformer-based semantic encoder.}
\label{fig2:encoder}
\vspace{-0.4cm}
\end{figure}
% [width=5in,height=2.3in]

\subsection{QoE Model and Problem Formulation}
Video transmission performance is closely related to user experience. In order to evaluate the user experience of video semantic communication, we propose a QoE model based on MIoU which is given by:
\begin{equation}
    {QoE}_{n}= \alpha {MIoU}_{n}- \beta {{T}_{n}}- |{{b}_{n}}-{{b}_{n-1}}|,
    \label{eq:6}
\end{equation}
where $\alpha$ and $\beta$ are hyperparameters, ${{b}_{n}}$ is the bitrate of video chunk $\boldsymbol{v}_{b_n}$ at the current moment, $|{{b}_{n}}-{{b}_{n-1}}|$ is the bitrate switching smoothness, and ${{T}_{n}}$ is the rebuffering time. Rebuffering, a result of exhausting data in the buffer area necessitating data reloading for continued playback, can significantly contribute to transmission delay. The aim of our ABRVSC scheme is to maximize QoE during video transmission over wireless networks, and the problem is formulated as follows,  
\begin{equation}
\begin{aligned}
& \underset{b_n}{ \max } \ \frac{1}{N} \sum_{n=1}^{N} {QoE}_{n} \\
& \text { s.t. } b_n \in \mathcal{B}.
\label{eq:7}
\end{aligned}
\end{equation}
where $N$ is the number of total video chunks. From (\ref{eq:6}), we can see that the QoE maximization problem~(\ref{eq:7}) involves MIoU, delay, and smoothness optimization. MIoU optimization relies on accurate semantic information extraction from the video, while delay and smoothness depend on proper bitrate selection based on accurate network condition prediction. To tackle these challenges, we proposed a semantic codec and an ABR algorithm which are explained in detail in Section III.

% Thus we use a rough function to describe the trend between them. Assuming that video bitrate is B, then the general functional form between them can be given by: 
% $$MIoU = a + b \ log B + c {{} \ (log B) ^ {2}} ,$$
% where $a, b, c $ are coefficients. In this functional form, when the bitrate is low, the MIoU will be low, because the details in the video may be lost; When the bitrate is high, $MIoU$ will gradually increase with the increase of bitrate, but the improvement range will gradually decrease.

\section{Semantic Codec and ABR Algorithm}
% Upon receiving a server request for a video, the device adaptively selects the appropriate bitrate for each video chunk based on the ABR algorithm, taking into account the current network conditions. 
% Having established the aim of maximizing the QoE for users in our ABRVSC system, we proceed to describe the key components responsible for achieving this goal， namely the MBSE module, the MBSD module, and the ABR algorithm. The design and implementation of these components are tailored to ensure the maximization of QoE, taking into account factors such as network conditions, transmission delay, and accurate semantic representation of video content.

With the aim of maximizing the QoE for the device in the ABRVSC scheme, we delve into the critical components responsible for realizing this objective, including the MBSE module, the MBSD module, and the ABR algorithm. The design and implementation of these components are adapted to optimize the QoE, considering the precise semantic representation of the video and ABR algorithm. In this section, we first introduce the MBSE and MBSD modules, which implement a multiple-bitrate video codec via deploying multiple compression ratios. Subsequently, we provide a detailed description of the video semantic communication ABR algorithm.

% The design and implementation of these components are adapted to optimize the QoE, considering the precise semantic representation of the video and ABR algorithm. 

\subsection{Semantic Encoder}
In order to extract the video semantic information, we adopt the scheme that processes video frames. The video frames are extracted using H.264 encoding standard (a widely adopted standard for video compression) and then fed into the semantic encoder. In the semantic encoder as shown in Fig.~\ref{fig2:encoder}, a swin transformer-based model \cite{liu2021swin} is employed to perform semantic segmentation on the input video frames, separating and
identifying various objects within each frame. Key semantic features, including object types, shapes, and locations, are extracted from the frame. Next, the semantic encoder encodes the extracted semantic features, generating a compact semantic representation, which is then transmitted over the wireless channel.

% These features not only maintain the essential information within the frame, but also create advantageous circumstances for the subsequent encoding and decoding processes.

% Our semantic encoder adopts a structure based on Swin Transformer model. Swin Transformer is a deep learning model for handling visual tasks. Compared to traditional convolutional neural networks (CNN), Swin Transformer uses a novel window moving mechanism to build a scalable architecture while also efficiently handling inputs of different scales\cite{liu2021swin}. 

% Fig.~\ref{fig2:encoder} shows the video segmentation model diagram, the feature dimension of the input video chunk is $ H\times W\times 3$, where $H$ and $W$ respectively represent the length and width of each frame of the input video chunk. The last dimension is 3, which is the number of channels is 3. 

% The key component of STBs is the Attention Mechanism: STB employs a shifted window-based self-attention mechanism that allows for efficient computation and improved scalability. This mechanism segment the input feature map into non-overlapping local windows and outputs self-attention within each window. By shifting the window positions between cascaded STBs, the STB captures long-range dependencies without significant increase in computational complexity.
\begin{figure}[t]
\centering
\includegraphics[width=1\linewidth]{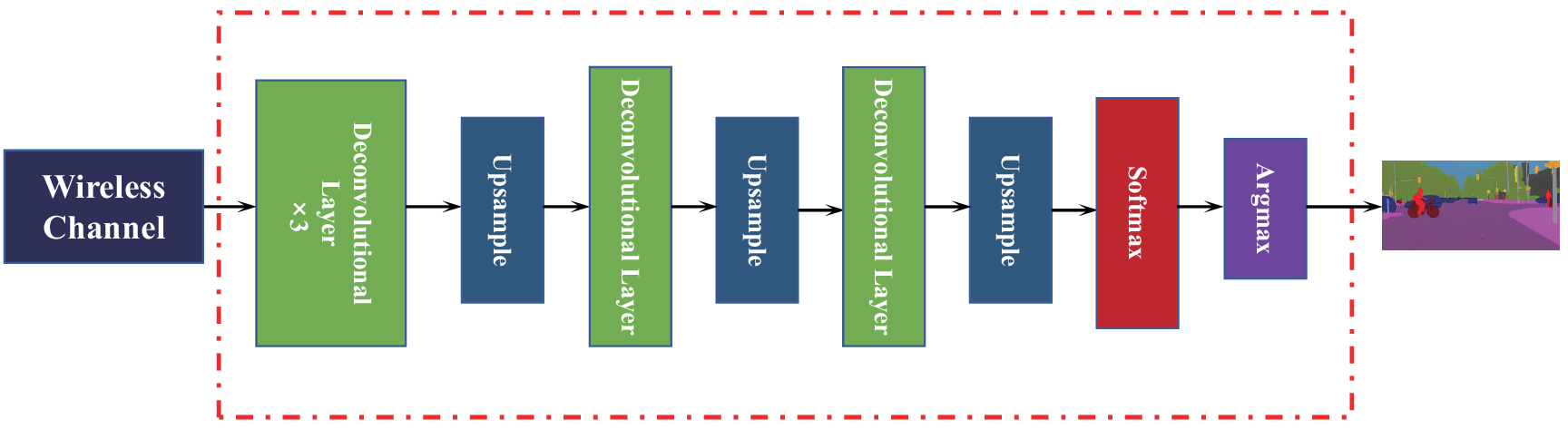}
\setlength{\abovecaptionskip}{-0.3cm}
\caption{The structure of convolutional neural network (CNN) based semantic decoder.}
\label{fig4:decoder}
\vspace{-0.5cm}
\end{figure}
% [width=2.9in]

The semantic encoder \cite{liu2021swin} is composed of a patch partition layer, four stages, a downsampling layer, and a convolutional layer. Each stage contains a common unit with several swin transformer Blocks (STBs) that can partition the input frame into non-overlapping local windows and conduct self-attention calculations within each local window. This self-attention mechanism enables STBs to allocate varying importance to distinct input elements while handling particular elements. Concurrently, STBs employ a hierarchical architecture with shifted windows, allowing the STBs to capture both local and global contextual information. These characteristics allow the model to selectively concentrate on particular regions of the input frames that are more pertinent to the objective of semantic segmentation, thereby enhancing its performance in this task. The self-attention \cite{liu2021swin} operation of the window can be given by:
\begin{equation}
    Attention(\boldsymbol{Q},\boldsymbol{K},\boldsymbol{V})=Softmax(\frac{\boldsymbol{Q}{{\boldsymbol{K}}^T}}{\sqrt{d}}+\tau)\boldsymbol{V},
\label{eq:8}
\end{equation}
where $\boldsymbol{Q}$, $\boldsymbol{K}$, and $\boldsymbol{V}$ are the corresponding query, key, and value matrices for the features of each window, respectively. In (\ref{eq:8}), $d$ is the dimension of the $\boldsymbol{K}$ matrix and $\tau$ is the relative position deviation. 

Following the downsampling layer, the features are passed through a convolutional layer consisting of $G$ filters, each with size 1 and stride 1. The video compression ratio of the semantic encoder can be controlled by adjusting the value of $G$. When the values for $G$ are 128, 64, 32, and 16, the corresponding compression ratios are 6, 12, 24, and 48, respectively.

\subsection{Semantic Decoder}

% The workflow of the proposed semantic decoder consists of three main steps: semantic decoding, image reconstruction, and video creation from frames. First, after receiving the transmitted semantic representation, the receiver reconstructs the semantic features of the original image. This step aims to ensure the maximal recovery of the semantic information of the original image whenever possible. Secondly, the original image is reconstructed based on the decoded semantic features. The objective of the image reconstruction is to minimize the loss of image quality while preserving the semantic information of the original image. This implies that the reconstructed image should accurately restore the boundaries and attributes of the objects in the original image, thereby achieving high-quality semantic segmentation results. Lastly, after reconstructing the individual images based on the decoded semantic features, these images are combined in their correct sequence to generate a video. The goal of this step is to seamlessly merge the reconstructed image frames to create a coherent and visually consistent video that maintains the overall flow and context of the original content.

% [width=2.9in]

In the semantic decoder, the receiver reconstructs semantic features from the transmitted representation, with the objective of achieving maximal semantic information recovery of the original frames. Then, the segmentation frames are reconstructed using decoded semantic features, minimizing frame quality loss while preserving semantic information. Finally, the reconstructed frames are combined into a video.

Fig.~\ref{fig4:decoder} presents the semantic decoder for video segmentation, which primarily comprises several deconvolutional layers, multiple upsampling layers, a softmax activation layer, and an argmax layer \cite{pan2022image}. The first three deconvolution layers aim to mitigate the effect of noise. Subsequently, in order to complete the details of the frame, three upsampling layers are interleaved between the deconvolutional layers, allowing the feature dimensions to gradually match the input in terms of length and width. After activation of the softmax activation layer, the obtained feature dimension is $H\times W\times c$, where $H$ and $W$ are respectively the height and width of each frame within the input video chunk, and $c$ is the number of categories for semantic segmentation. The value of the last dimension signifies the prediction probability of pixel multi-class classification. Lastly, semantic segmentations of video frames are acquired through the argmax operation with a dimension of $H\times W$.

% and the value of last dimension corresponds to the category subscript with the highest prediction probability value. 

% After activation through an softmax activation layer, the feature dimension we get is the semantic feature of $H\times W\times N$, where $N$ is the number of categories of semantic segmentation, and the value of the last dimension is the prediction probability of pixel multi-category classification. Finally, we perform the Argmax operation to obtain the semantic segmentation of the video chunk image frame. Its characteristic dimension is $H\times W\times 1$, and the value of the last dimension is the category subscript with the largest prediction probability value.
The objective of the decoder is to accurately classify each pixel in a frame. Therefore, we utilize the cross entropy of the multi-class classification for each pixel as the loss function. For a batch of frames, the loss function of the entire system can be given by:
% \begin{equation}
%     {Loss}=\frac{-\sum\limits_{s=1}^{S}{\sum\limits_{p=1}^{H\times W}{\sum\limits_{i=0}^{c-1}{({{p}_{i}}\log ({{{\hat{p}}}_{i}}))}}}}{S\times H\times W},
% \end{equation}

% \begin{equation}
% {Loss}=\frac{-\sum\limits_{s=1}^{S}{\sum\limits_{p=1}^{H\times W}{\sum\limits_{i=0}^{c-1}{\sum\limits_{j=0}^{c-1}{(\frac{{{p}_{ij}}}{\sum\limits_{k=0}^{c-1}{{p}_{ik}}}\log(\frac{{{p}_{ij}}}{\sum\limits_{k=0}^{c-1}{{p}_{kj}}}))}}}}}{S\times H\times W},
% \end{equation}
\begin{equation}
{Loss}=\frac{-\sum\limits_{s=1}^{S}{\sum\limits_{l=1}^{H\times W}{\sum\limits_{i=0}^{c-1}{({{p}_{s, l, i}}\log ({{{\hat{p}}}_{s, l, i}}))}}}}{S\times H\times W},
\end{equation}
where $S$ is the batch size, ${{{\hat{p}}}_{s, l, i}}$ is the predicted probability that the pixel at position $l$ in the $s$ frame is classified into category $i$, and the value of ${{p}_{s, l, i}} \in \{0,1\}$ is the label of the frame pixel at position $l$ in the $s$ frame that belongs to category $i$.
% where $S$ is the batch size, ${{\hat{p}}_{i}}$ is the probability that the pixel is classified into category $i$, and the value of ${{p}_{i}} \in \{0,1\}$ is the actual label of the frame pixel. Given the loss function, an end-to-end approach is utilized to train the codec. 

% Consequently, the semantic decoder is capable of mitigating the channel influence when decoding the received semantic features.

\begin{figure}[t]
\centering
\vspace{-0.2cm}
\includegraphics[width=0.7\linewidth]{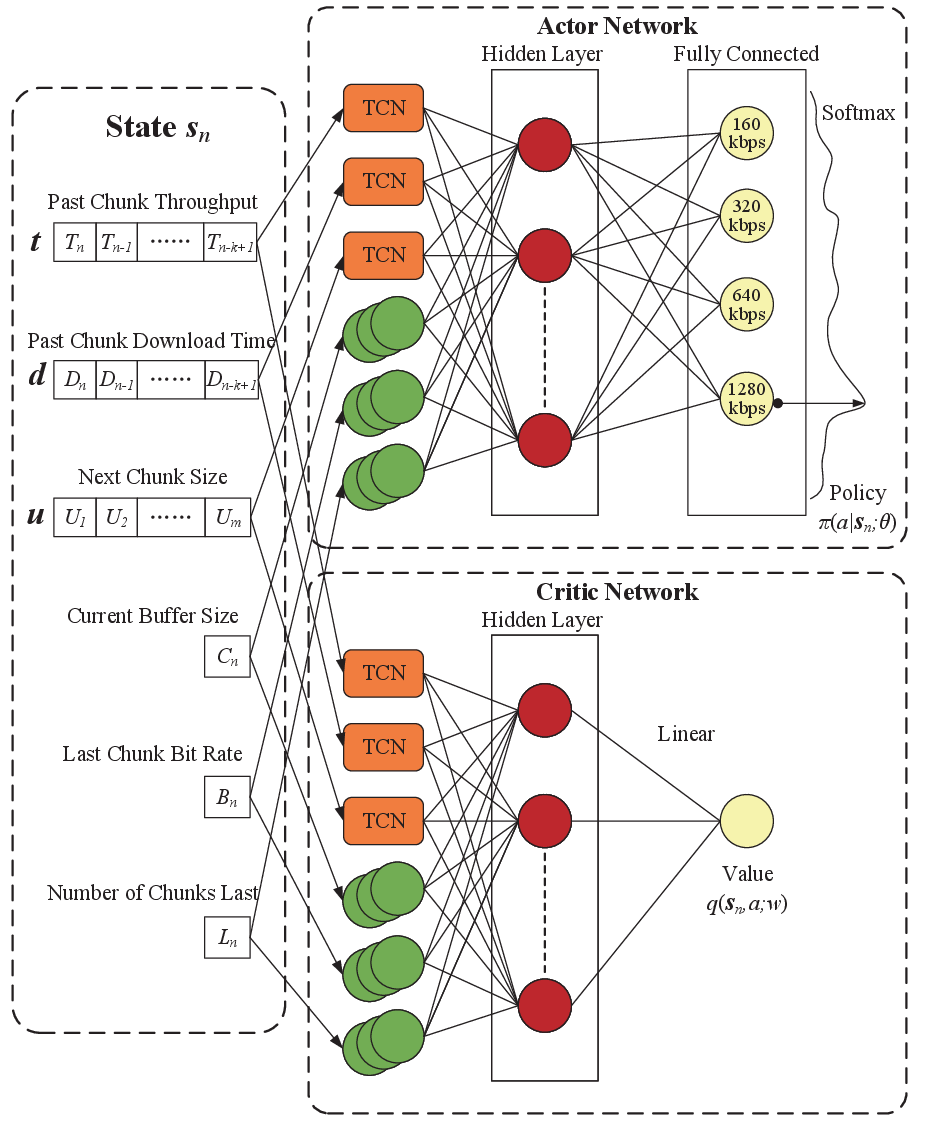}
\setlength{\abovecaptionskip}{-0.1cm}
\caption{The actor-critic based ABR model.}
\label{fig6:model}
\vspace{-0.5cm}
\end{figure}

\subsection{Actor-Critic based Video ABR Algorithm}
To guarantee both low transmission delay and high semantic accuracy under various network conditions, the device must select an appropriate bitrate based on the video ABR algorithm. Consequently, we employ Actor-Critic reinforcement learning (RL) for ABR, which efficiently converges by leveraging temporal difference error.
% \begin{figure}[htb]
% \centering
% \includegraphics[width=1\linewidth]{./picture/system_model/Actor.eps}
% \caption{Actor-Critic}
% \label{fig5:actor}
% \vspace{-0.4cm}
% \end{figure}
The Actor-Critic based ABR model is shown in Fig.~\ref{fig6:model}. In the Actor-Critic algorithm, there are two main components: Actor network and Critic network. For each video chunk, state $\boldsymbol{s}_{n}$, which indicates the communication network condition, is input into the neural network of an agent. The RL agent deployed on the device selects the transmission bitrate $b_n$ as the action $a_n$. The agent then executes the action, observes the reward from environmental feedback, and the RL environment evolves to the next. After the action is performed, the Critic estimates the state value based on the action $a_n$ and network state $\boldsymbol{s}_{n}$, providing the Actor with an evaluation of the action, denoted by value $q_n$. The Actor then updates its model parameters according to $q_n$.

\begin{algorithm}[t]
 \renewcommand{\algorithmicrequire}{\textbf{Input:}}
 \renewcommand{\algorithmicensure}{\textbf{Output:}}
 \caption{\small {Video Bitrate Adaptive Algorithm based on Deep Reinforcement Learning.}}
 
 \label{alg:FL_comb}
 \begin{algorithmic}[1]
%   \REQUIRE latent dimension $K$, $G$, target predicate $p$
%   \ENSURE $U^{p}$, $V^{p}$, $b^{p}$
%   \STATE \textbf{Initialize:} Randomly initialize parameters $\boldsymbol{\theta}^{(0)}$ and $\boldsymbol{\varphi}^{(0)}$， $i = 0$. 
    \small
        \STATE \textbf{Initialization:} Initialize policy network parameter $\theta $ and value network parameter $w$. Set the hyperparameter $\alpha$, $\beta$ and discounting factor $\gamma$. Initialize the environment and get the initial state ${\boldsymbol{s}_{0}}$. Select action $a_0$ according to the policy network $\pi (a|{\boldsymbol{s}_{0}};\theta )$.
     \STATE \textbf{while} training epochs $<$ Total training epochs $M$ \textbf{do}:       
     \STATE \quad \textbf{while} video chunks $<$ Total video chunks $N$ \textbf{do}:
    % \STATE \quad \textbf{for} local model $\boldsymbol{w_{i}}$ of $U$ users, \textbf{do}:
    \STATE \quad \quad Observe the network state ${\boldsymbol{s}_{n}}$ and randomly sample the \\ \quad \quad bitrate ${{a}_{n}}$ according to the policy function $\pi (\cdot |{\boldsymbol{s}_{n}};{{\theta }_{n}})$.
       \STATE \quad \quad Play the video chunk with bitrate ${{a}_{n}}$ and get the new \\ \quad \quad network state ${\boldsymbol{s}_{n+1}}$ and reward ${{r}_{n}}$. 
       % \STATE \quad \quad Calculate value network ${{q}_{t}}=q({{s}_{t}},{{a}_{t}};{{w}_{t}})$ and \\ \quad \quad ${{q}_{t+1}}=q({{s}_{t+1}},{{\tilde{a}}_{t+1}};{{w}_{t}})$.

        \STATE \quad \quad ${{\delta }_{n}}={{q}_{n}}-({{r}_{n}}+\gamma \cdot {{q}_{n+1}})$ $\hfill\triangleright$ calculate TD error 
        \STATE \quad \quad  ${{d}_{w,n}}=\frac{\partial q({\boldsymbol{s}_{n}},{{a}_{n}};w)}{\partial w}{{|}_{w={{w}_{n}}}}$   $\hfill\triangleright$ calculate derivative
        \STATE \quad  \quad ${{w}_{n+1}}\leftarrow {{w}_{n}}-\varsigma \cdot {{\delta }_{n}}\cdot {{d}_{w,n}}$  $\hfill\triangleright$ update parameter $w$
        \STATE \quad \quad  ${{d}_{\theta ,n}}=\frac{\partial \log \pi ({{a}_{n}}|{\boldsymbol{s}_{n}},\theta )}{\partial \theta }{{|}_{\theta ={{\theta }_{n}}}}$ $\hfill\triangleright$ calculate derivative
        \STATE \quad  \quad ${{\theta }_{n+1}}\leftarrow {{\theta }_{n}}+\sigma \cdot {{q}_{n}}\cdot {{d}_{\theta ,n}}$   $\hfill\triangleright$ update parameter $\theta$
        \STATE  \quad \textbf{end while} 
        \STATE  \textbf{end while}
        \STATE \textbf{Output:}  Final policy network parameter ${{\theta }^{*}}$, value network parameter $w^{*}$ and reward curve.
        % \STATE \textbf{Output:} 
     \end{algorithmic}  
     \vspace{-0.1cm}
    % \vspace{-0.2cm}
\end{algorithm}

% Through continuous training of the model, Actor will obtain higher and higher value assessment $q$ given by Critic. At the same time, the video playing terminal will generate a reward value $r$ and feed back to Critic, and update its own model parameters according to the feedback information. Through continuous training of the model, the value assessment $q$ of the actions of object Actor will become more and more accurate. The final video bitrate decision will be made in the direction of increasing the reward value $r$.

For each video chunk, the input state space \cite{mao2017neural} consists of six parameters: $\boldsymbol{s}_{n}=({\boldsymbol{t}},{\boldsymbol{d}},\boldsymbol{u},{{C}_{n}},{{B}_{n}},{{L}_{n}})$. ${\boldsymbol{t}}$ is the throughput vector for the past $k$ video chunks, ${\boldsymbol{d}}$ is download time vector for the past $k$ video chunks, and $\boldsymbol{u}$ is a vector containing $m$ available sizes for the next video chunk. They are processed by a Temporal Convolutional Network (TCN) layer with 64 filters, kernel size 3, and stride 1. Here, we employ the TCN due to its superior performance in processing time series data. This is largely attributed to its unique utilization of causal convolution, dilated convolution, and residual operations. The current buffer occupation ${{C}_{n}}$, the last video chunk bitrate ${{B}_{n}}$, and the number of remaining video chunks ${{L}_{n}}$ are extracted by a fully connected neural network with 128 neurons. The output features of these layers are then transmitted to a hidden layer of 128 neurons. The hidden layer is fully connected to an output layer which uses the softmax function to select the action ${{a}_{n}}$. 
\begin{figure*}[t]
	\setlength{\abovecaptionskip}{-5pt}
	\setlength{\belowcaptionskip}{-10pt}
	\centering
	\begin{minipage}[t]{0.33\linewidth}
		\centering
		\includegraphics[width=2.5in]{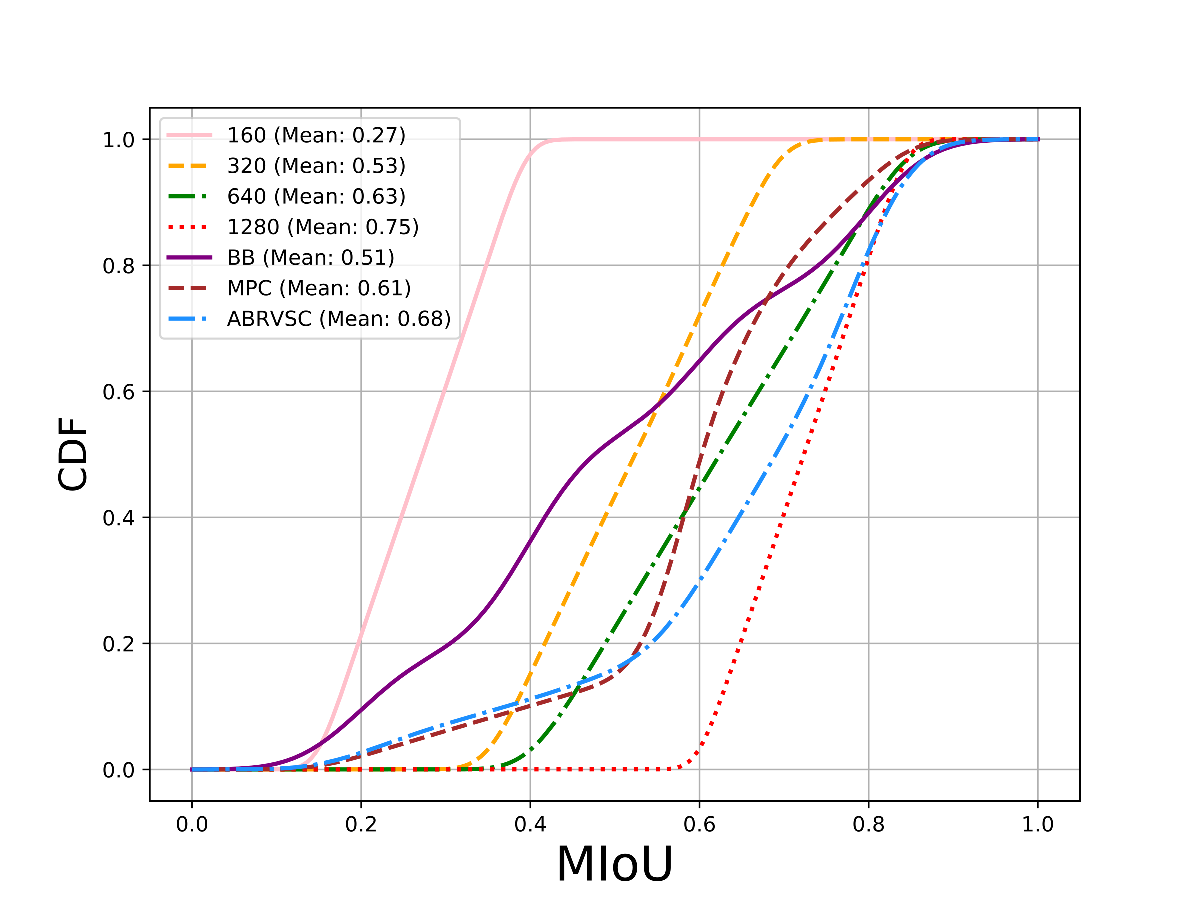}
		\caption{CDF of MIoU.}
		\label{fig7:res1}
	\end{minipage}%
	\begin{minipage}[t]{0.33\linewidth}
		\centering
		\includegraphics[width=2.5in]{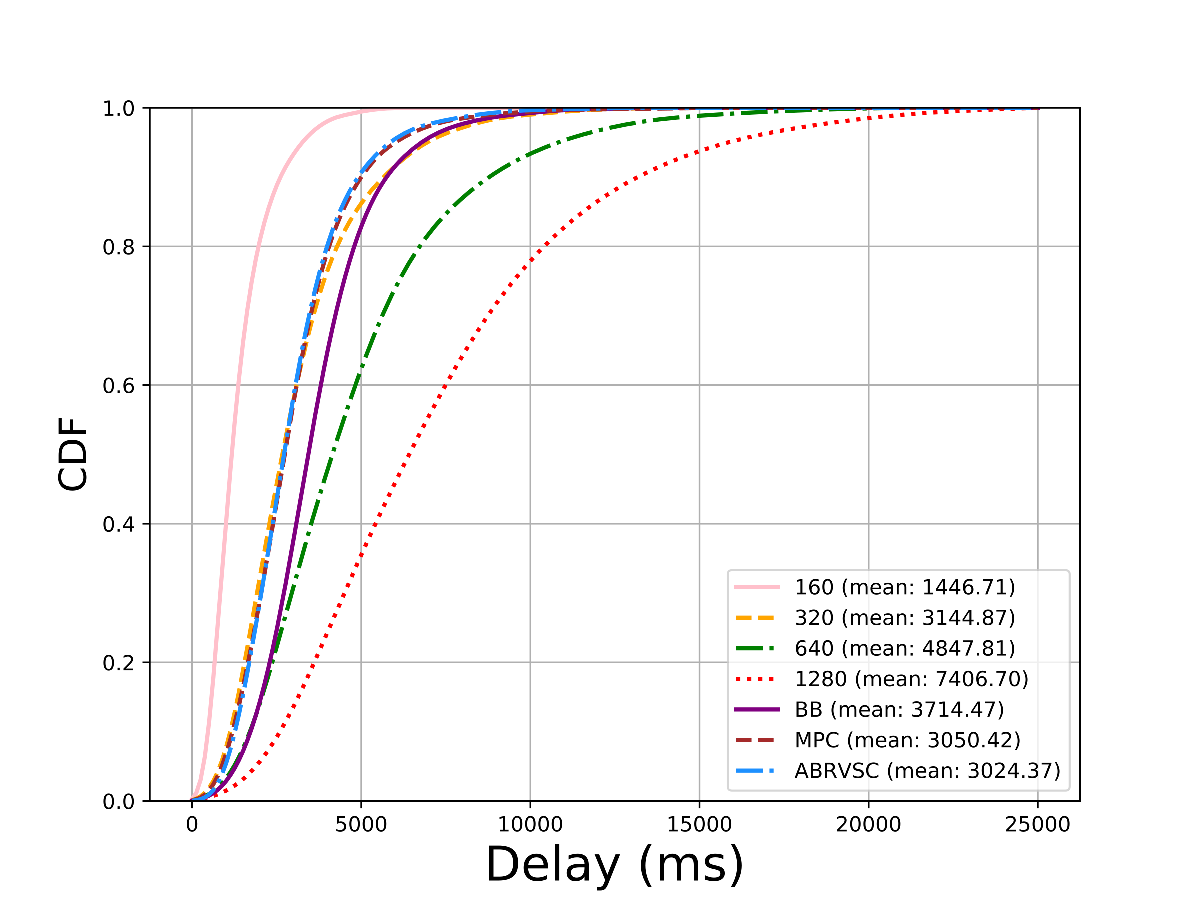}
		\caption{CDF of delay.}
		\label{fig8:res2}
	\end{minipage}
	\begin{minipage}[t]{0.33\linewidth}
		\centering
		\includegraphics[width=2.5in]{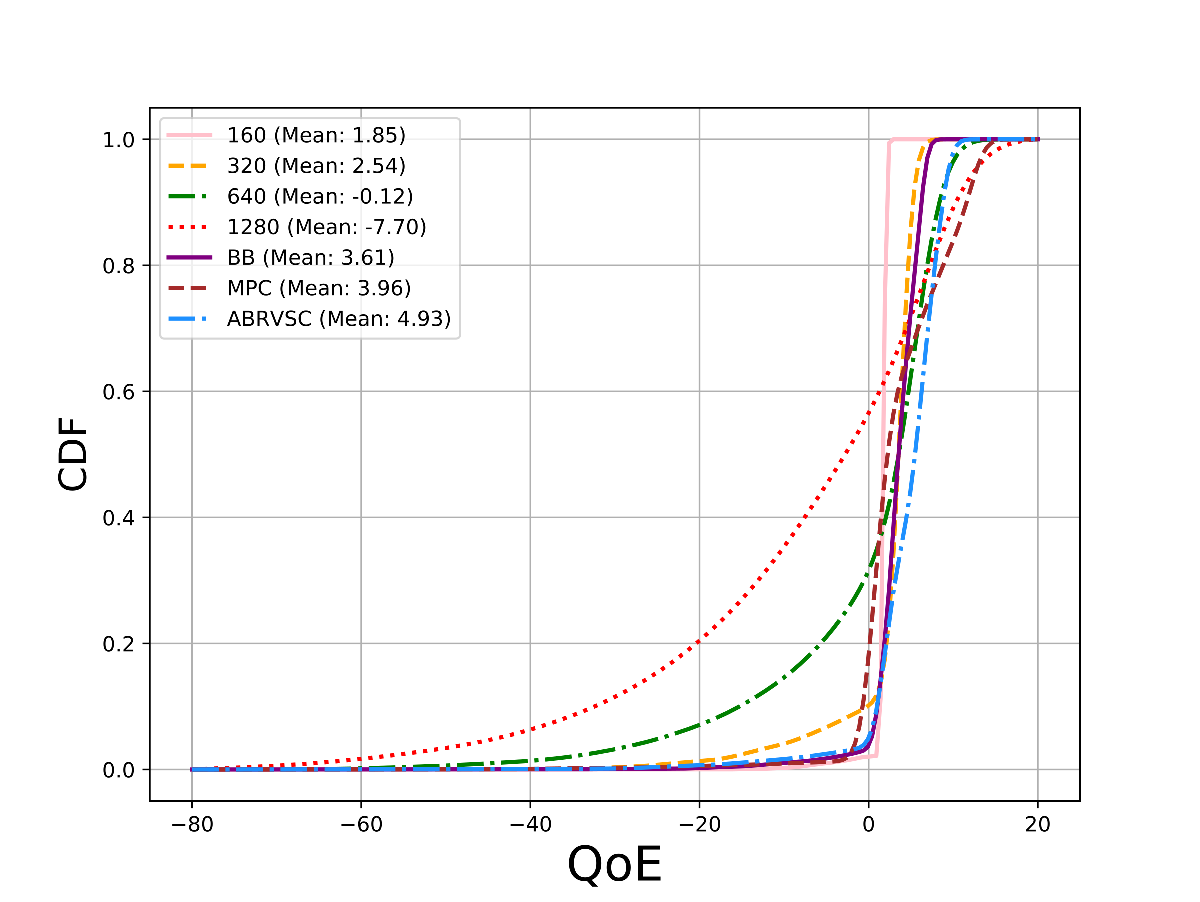}
		\caption{CDF of QoE.}
		\label{fig9:res3}
	\end{minipage}
\vspace{-0.7cm}
\end{figure*}
Different from the Actor, the Critic employs a linear activation at the output layer to evaluate the QoE of the video service. The training objective is to maximize the reward value ${{r}_{n}}$, which corresponds to the QoE introduced in Section II. The reward function is given by:
\begin{equation}
    {{r}_{n}}=\alpha {MIoU}_{n}- \beta {{T}_{n}}- |{{b}_{n}}-{{b}_{n-1}}|.
\end{equation}

We use ${{V}_{\pi }}({\boldsymbol{s}_{n}})=\sum\limits_{a}{\pi (a|{\boldsymbol{s}_{n}}){{Q}_{\pi }}({\boldsymbol{s}_{n}},a)}$ as the state value function, representing the average value of all actions. Then, ${{V}_{\pi }}({\boldsymbol{s}_{n}})$ can be approximated by a neural network weighted with $w$, which is given by:
\begin{equation}
    {{V}_{\pi }}({\boldsymbol{s}_{n}}) \approx V({\boldsymbol{s}_{n}};\theta ,w)=\sum\limits_{a}{\pi (a|{\boldsymbol{s}_{n}};\theta )} q({\boldsymbol{s}_{n}},a;w).
\end{equation}
The value network (Critic network) $q({\boldsymbol{s}_{n}},a;w)$ is updated based on $V({\boldsymbol{s}_{n}};\theta ,w)$ to enhance the scoring accuracy, which in turn allows for better estimation of the sum of future expected rewards, bringing $V({\boldsymbol{s}_{n}};\theta ,w)$ closer to the actual average value. The policy network (Actor network) $\pi (a|{\boldsymbol{s}_{n}};\theta )$ is updated according to $q({\boldsymbol{s}_{n}},a;w)$ in order to increase the state value $V({\boldsymbol{s}_{n}};\theta ,w)$. The complete Actor-Critic based ABR algorithm is presented in Algorithm 1, which leverages Temporal Difference (TD) error to guide both the value function estimation and the action selection optimization. 

Moreover, to address the parallel training problem of Actor-Critic in multi-core or distributed computing environments, we adopt \cite{mnih2016asynchronous} the Asynchronous Advantage Actor-Critic (A3C) algorithm. This approach employs multiple agents with each possessing its own Actor and Critic. These agents train asynchronously and subsequently aggregate their experiences into a global network.

\section{Simulation Results}

% This section first introduces the experimental environment configuration, experimental dataset, experimental parameter setting, and baseline selected in this paper. After that, the experimental results will be shown and the performance will be analyzed compared with the baselines.

% \subsection{Simulation environment}
In our simulations, the device has a 60 seconds video playback buffer and employs four different bitrates: $\mathcal{B}$ = \{160 kbps, 320 kbps, 640 kbps, 1280 kbps\}. The specific experimental parameter settings are presented in Table I. To evaluate the performance of our algorithm in a real network environment, we adopt a mixed dataset of FCC and HSDPA \cite{riiser2013commute} as our network bandwidth dataset and CamVid as the video dataset \cite{brostow2009semantic}, respectively. In the network bandwidth dataset, we utilize 75\% dataset for training the model, and 25\% for testing. For the video dataset, the dataset is partitioned into 367 frames for training, 101 for validation, and 233 for testing. We compare our algorithm with a) OCRNet+JPEG (semantic segmentation model with JPEG coding), b) Buffer-Based (BB) Algorithm \cite{huang2014buffer}, c) Model Predictive Control (MPC) Algorithm \cite{yin2015control}, and d) Fixed bitrate transmission schemes, which bitrates $\in \mathcal{B}$.

\begin{table}[t]
    \centering
    \caption{Simulation Parameters.}
    \vspace{-0.1cm}
        \begin{tabular}{|>{\centering\arraybackslash}p{1cm}|*{6}{>{\centering\arraybackslash}p{0.5cm}|}>{\centering\arraybackslash}p{0.8cm}|}
        \hline
        Parameter & \multicolumn{1}{>{\centering\arraybackslash}p{1.0cm}|}{$\sigma$} & $\varsigma$ & $\alpha$ & $\beta$ & k & m & c\\
        \hline
        Value & \multicolumn{1}{>{\centering\arraybackslash}p{1.0cm}|}{0.0001} & 0.001 & 9.6 & 4.3 & 8 & 4 & 32\\
        \hline
        \end{tabular}
    \label{tab:simu_parameter}
    \vspace{-0.7cm}
\end{table}

In Fig.~\ref{fig7:res1}, we show the cumulative distribution function (CDF) of MIoU for all considered schemes with different network conditions. The CDF indicates the probability distribution of a random variable. From Fig.~\ref{fig7:res1}, we see that although MIoU of the ABRVSC scheme is slightly lower than that of 1280 kbps scheme, the ABRVSC scheme can achieve 151.85\%, 28.30\%, 7.94\%, 33.33\%, and 11.48\% gain in terms of MIoU compared to 160 kbps, 320 kbps, 640 kbps, BB and MPC schemes, respectively. This is due to the fact that the 1280 kbps scheme pursues high semantic accuracy at the cost of delay caused by huge transmission data amount, while the Actor-Critic based ABR algorithm enables the ABRVSC scheme to select high bitrates in the case of high bandwidth, thus improving MIoU. Fig.~\ref{fig7:res1} demonstrates that the ABRVSC scheme can attain high MIoU for video semantic communication.

% \begin{figure}[!htbp]
% \centering
% \includegraphics[width=0.9\linewidth]{./picture/36000/CDF_reward.eps}
% \caption{CDF of total rewards}
% \label{fig8:res2}
% \vspace{-0.2cm}
% \end{figure}'

% % [width=2.9in]

Fig.~\ref{fig8:res2} shows the CDF of transmission delay for all considered schemes with different network traces. From Fig.~\ref{fig8:res2}, we see that although the average transmission delay of the ABRVSC scheme is higher than the 160 kbps scheme, the ABRVSC scheme can reduce 3.83\%, 37.61\%, 59.17\%, 18.58\%, and 0.85\% average transmission delay compared to 320 kbps, 640 kbps, 1280 kbps, BB and MPC schemes, respectively. This is because that the 160 kbps scheme at the cost of a lot of semantic accuracy in exchange for low transmission delay, while the accuracy of the network condition prediction enables the ABRVSC scheme to select low bitrates in the case of low bandwidth, thus reducing transmission delay. Fig.~\ref{fig8:res2} reflects the outperformance of ABRVSC scheme on guaranteeing low video service transmission delay.

% \begin{figure}[!htbp]
% \centering
% \includegraphics[width=0.9\linewidth]{./picture/next/bitrate.eps}
% \caption{CDF of total delays}
% \label{fig9:res3}
% \vspace{-0.2cm}
% \end{figure}
\begin{figure}[t]
\centering
\vspace{-0.15cm}
\includegraphics[width=0.7\linewidth]{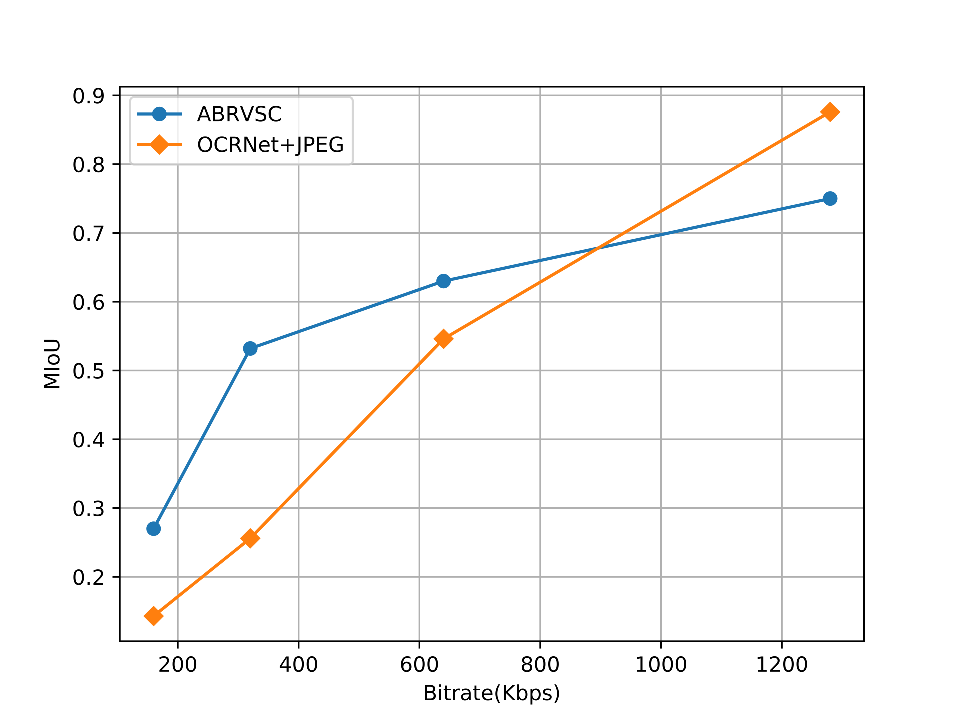}
\setlength{\abovecaptionskip}{-0.2cm}
\caption{MIoU of different models under different bitrates.}
\label{fig10:res4}
\vspace{-0.55cm}
\end{figure}
In Fig.~\ref{fig9:res3}, we show the CDF of QoE for all considered schemes with different network conditions. From Fig.~\ref{fig9:res3}, we see that the ABRVSC scheme can achieve 166.49\%, 94.09\%, 36.57\%, and 24.49\% gain of the QoE compared to 160 kbps, 320 kbps, BB and MPC schemes, respectively. This is due to the fact that the ABRVSC scheme can adjust the bitrate in time according to the real-time channel state to maximize the utilization efficiency of the channel resources, so as to obtain better QoE. Fig.~\ref{fig9:res3} demonstrates that the ABRVSC scheme outperforms the baselines in terms of QoE.

Fig.~\ref{fig10:res4} shows how the MIoU of ABRVSC scheme and traditional coding scheme change as the bitrate varies. From Fig.~\ref{fig10:res4}, we see that under low bitrates, the MIoU of the ABRVSC scheme surpasses that of the traditional scheme. Although at the bitrate of 1280 kbps, the ABRVSC scheme has a slightly lower MIoU compared to the traditional scheme, when the bitrate is 160 kbps, 320 kbps, and 640 kbps, the MIoU of the ABRVSC scheme is nearly 92.86\%, 107.03\%, and 15.38\% higher than that of the traditional scheme, respectively. \color{black} This is due to the fact that traditional schemes extract all the details of the video frame, and completely reconstruct the encoded frame under a high bitrate while the ABRVSC scheme can effectively extract the semantic information of video frames thus achieving better performance with low bitrates. Fig.~\ref{fig10:res4} demonstrates that the proposed ABRVSC scheme can achieve higher semantic accuracy under low bitrates compare to traditional coding scheme.

\vspace{-0.3cm}
\section{Conclusion}
In this paper, we proposed an ABRVSC system and first defined the QoE for video semantic communication and formulated a QoE maximization problem. To solve problem, we then proposed a swin transformer-based semantic codec and introduced an Actor-Critic based ABR algorithm to the semantic codec. 
Simulation results demonstrates that at low bitrates, the MIoU of the proposed ABRVSC scheme is nearly twice that of the traditional scheme. Moreover, the proposed ABRVSC scheme increases the QoE in video semantic communication, which exhibits more robustness against network variations.

\def\baselinestretch{0.84}
\bibliographystyle{IEEEtran}
\bibliography{IEEEabrv,ref}

\end{CJK}
\end{document}